\documentclass[10pt,conference]{IEEEtran}
\usepackage[utf8]{inputenc}
\usepackage[T1]{fontenc}
\usepackage{graphicx}
\usepackage{amsmath,amssymb}
\usepackage{amsthm}
\theoremstyle{definition}
\newtheorem{example}{Example}
\usepackage[hidelinks]{hyperref}
\usepackage{booktabs}
\usepackage{microtype}

\usepackage{tikz}
\usetikzlibrary{arrows,shapes,positioning}
\usepackage{float}
\begin{document}

\title{Regret Dominates Surprise: Design-Time Requirements Engineering for Agentic-AI Safety}

\author{
\IEEEauthorblockN{Nuwayyir Almohammadi}
\IEEEauthorblockA{\textit{School of Computer Science} \\
\textit{University of Birmingham}\\
Birmingham, United Kingdom \\
Nra406@student.bham.ac.uk}
\and
\IEEEauthorblockN{Rami Bahsoon}
\IEEEauthorblockA{\textit{School of Computer Science} \\
\textit{University of Birmingham}\\
Birmingham, United Kingdom \\
r.bahsoon@bham.ac.uk}
\and
\IEEEauthorblockN{Tao Chen}
\IEEEauthorblockA{\textit{School of Computer Science} \\
\textit{University of Birmingham}\\
Birmingham, United Kingdom \\
t.chen@bham.ac.uk}
}

\maketitle

\begin{abstract}
Requirements Engineers for Agentic-AI domains 
face challenges in evaluating and verifying the modeling, elaboration, and operationalization of safe autonomy in these systems. Mainstream frameworks, such as Goal-Oriented Requirements Engineering (\textsc{gore}), lack mechanisms to systematically address these challenges in the face of epistemic uncertainty.

We contribute to an approach that builds on \textsc{gore} to model and simulate safe autonomy in agentic-AI systems. We introduce a novel Regret-Dominance Mechanism(MS-RGR) to assist requirements engineers in operationalising for safer autonomy. MS-RGR uses two signals: epistemic surprise (novelty detection) and cognitive regret (evaluative risk) to model and simulate, the problem we coin as the trilemma problem for safe autonomy:
should the agent operate in routine autonomy, undergo reflective reasoning or escalate to human? 

We instantiate the \textsc{ms-rgr} in two workflow domains: elderly-care monitoring and autonomous driving decision sequences. A domain-agnostic 100-seed stochastic simulation with structural ablations shows MS-RGR reduces silent-failure incidence to near-zero and detects risk roughly 17.5× faster than a sensor-only baseline, while remaining formally traceable via LTL safety properties.
A retrospective proxy instantiation, applying the \textsc{dri}
gate post-hoc over recorded execution traces from 208
\textsc{AgentHarm} scenarios across seven \textsc{llm}s, shows the gate meaningfully improves harmful-task refusal only for the two models with strong baseline safety behavior (>80\% pre-gate refusal), e.g. 84.1\%→90.9\% for one model, while leaving the remaining five unaffected, indicating MS-RGR amplifies rather than substitutes for model-level safety training. We discuss this and other threats to validity, positioning MS-RGR as initial feasibility evidence for design-time, evaluative safety constraints in agentic-AI requirements engineering.
\end{abstract}


\section{Introduction}

Agentic-AI systems are increasingly being deployed as autonomous
decision-makers in high-stakes domains such as healthcare, banking,
autonomous driving, and child-facing applications, where misbehavior
or unsafe actions can cause real-world harm without immediate human
oversight~\cite{Amodei2016ConcreteSafety,
Andriushchenko2025AgentHarm:Agents}.
The shift towards Agentic-AI autonomy challenges requirements modeling,
elaboration, operationalization, and design-time assumptions, to account for the
`unknown unknowns' encountered in agentic environments and the design for safe autonomy, covering: routine autonomy with assured safety; undergoing reflective reasoning for safer autonomy; or overriding autonomy with human-in-the-loop interactive support for safety. To address this problem, we propose a multi-signal mechanism
that leverages \textsc{gore} to capture and operationalize the
interplay between surprise (epistemic mismatch) \cite{Samin2025SurpriseAdapt} and regret
(evaluative risk) in the modeling for safe autonomy in agentic-AI systems.

Consider a medical-monitoring agent trained on a standard hospital
data that encounters a rare, non-harmful cardiac irregularity.
Without an explicit ``unknown'' category, it misclassifies the
condition as cardiac arrest. Unlike traditional diagnostic AI systems, which produce a classification for a clinician to act upon, Agentic-AI systems possess autonomous executive authority to independently plan and execute a critical response \cite{Sapkota2026AIChallenges}, and administer a high-voltage
shock~\cite{Amodei2016ConcreteSafety} without immediate human oversight.
The functional requirement is satisfied; the evaluative
question---``would notifying a human caregiver have been
safer?''---was never asked.
This \emph{confident failure}~\cite{Amodei2016ConcreteSafety}
arises not from probabilistic uncertainty about outcomes but from
the absence of a mechanism to compare the safety of available
tactics at the moment of decision.
Although recent research has explored ``regret'' in \textsc{llm}
agents~\cite{Park2025DoGames} and the role of ``surprise'' as a
meta-cognitive signal for identifying model
incompleteness~\cite{Samin2025SurpriseAdapt,
Ghanmi2025EvaluatingSurprises}, these approaches remain largely
post-hoc or runtime-focused  which may prove insufficient for safety-critical contexts. Both tend to operate after the decision cycle has begun, offering limited protection when interventions are instantaneous and irreversible. More critically, relying on such signals risks leaving boundary decisions, such as when to override agent autonomy and escalate to a human, to emergent runtime heuristics rather than verifiable engineering constraints, with no clear guarantee that escalation occurs before harm is done.
This work argues that Requirements Engineering plays a vital role
in proactively modeling verifiable safety constraints that mitigate
the risk of confident failures characteristic of the agentic
era~\cite{Amodei2016ConcreteSafety}. We operationalize regret
and surprise at design time, and provide feasibility evidence through stochastic simulation and proxy instantiation against AGENTHARM benchmark. AGENTHARM is a set of 208 agentic evaluation scenarios spanning eight harm categories, where harmful intent is embedded within tool-use framing rather than stated explicitly, making it a suitable stress-test for design-time safety gating mechanisms such as MS-RGR for the said purpose.

Following Meta-cognitive Explanation-Based (\textsc{meb})
theory~\cite{Foster2015WhyDifficulty}, we frame \emph{surprise} as
a signal of explanatory difficulty~\cite{Foster2015WhyDifficulty,
Ghanmi2025EvaluatingSurprises} that identifies mismatches between
runtime outcomes and the agent's internal model.
This multi-signal approach enables a transition from routine
execution to a regret-aware evaluation
phase~\cite{Samin2025SurpriseAdapt}, where design-time constraints
govern the selection of resolution tactics.
By anchoring these cognitive signals in a \textsc{gore} framework,
we provide a structured method to define safety thresholds that
remain traceable to stakeholder requirements.

\noindent\textbf{Contributions.} We contribute to an approach that requirements engineers can leverage to model and simulate the trilemma problem of safe autonomy in agentic-AI systems, considering: routine autonomy with guaranteed safety; undergoing reflective reasoning for safer and "cautioned" autonomy; or overriding autonomy with human-in-the-loop interactivity for safety.
We build on van Lamsweerde and Letier's goal-obstacle modeling and resolution strategies~\cite{VanLamsweerde2000HandlingEngineering} and \textsc{mape-k} architecture of Cailliau and van Lamsweerde~\cite{Cailliau2017RuntimeGoals} as engineering substrate.  We introduce a novel regret-dominance mechanism that complements safety probabilistic satisfaction metrics with ordinal and design-time thresholds to address the trilemma. The mechanism makes novel use of cognitive regret as a \emph{design-time} safety signal when handling  \textsc{gore} modeling and obstacles resolution. It shifts evaluative safety decisions from aftermath runtime heuristics to a priori verifiable safety thresholds specified, simulated and verified before deployment. Grounded in Simon's \textit{Satisficing} principle~\cite{Simon1955REPRINTED:Man}, the framework treats mathematically optimal execution as a system failure when it provokes excessive user regret, enforcing structured halts within predefined engineering bounds. Specifically, we make the following interrelated major contributions to Requirements Engineering of Agentic-AI Safety:

\begin{itemize}

    \item \textbf{A Multi-Signal Gating Mechanism (Regret-Dominance Gate):} We design a novel dynamic traffic-control mechanism that monitors autonomous runtime behavior by consolidating two unique trace metrics: the \textit{Epistemic Surprise Signal} and the \textit{Evaluative Regret}. The resulting Regret-Dominance Gate adjudicates 
the autonomy trilemma across three cognitive tiers:
\begin{itemize}
    \item \textit{System 1 (Routine Autonomy):} 
    Safe automated execution under low regret.
    \item \textit{System 2 (Reflective Reasoning):} 
    Regret-aware adjustment applying \textit{Loss Aversion}~\cite{Kahneman1979ProspectRisk,Kahneman2011ThinkingSlow,Kahneman2002MAPSCHOICE} 
    when $\text{DRI}\!\geq\!\tau$.
    \item \textit{Human Escalation:} 
    Agent replaced by human when 
    $\text{DRI}\!\geq\!\tau_{\text{crit}}$.
\end{itemize}
    
    \item \textbf{Pre-Modeled Design-Time Resolution Tactics:} We operationalize runtime adaptation by enabling requirements engineers to embed automated, deterministic resolution tactics directly into the goal model based on risk severity. Moderate risks trigger goal weakening via the \textit{Back-off} tactic, while critical risks trigger \textit{Goal Sacrificing}, dropping the functional goal entirely to initiate an emergency escalation that replaces the software agent with a human caregiver. Crucially, these adaptations represent explicit design-time RE boundary decisions rather than autonomous runtime model choices, thereby preserving a safe, cooperative relationship with the user.
\end{itemize}
\href{https://doi.org/10.5281/zenodo.20928801}{Anonymised replication package available here}.

The primary users of \textsc{ms-rgr} are requirements engineers evaluating safety requirements and their constraints before deployment.
The Regret-Dominance Gate governs agentic behavior under epistemic
uncertainty by constraining tactic selection within design-time
safety thresholds; human users benefit indirectly through
regret-aware, \textbf{evaluatively safeguarded} agentic behavior under
epistemic uncertainty.\textsc{ms-rgr} reduces silent-failure to 
near-zero vs.\ 96\% sensor-only baseline 
and detects risk 17.5$\times$ faster 
(Section~\ref{sec:eval}).

\section{Background and Motivation}
We operationalize surprise--regret  distinction using two complementary
cognitive constructs.
\emph{Surprise} is modeled using \textsc{meb}
theory~\cite{Ghanmi2025EvaluatingSurprises,
Foster2015WhyDifficulty}: a meta-cognitive signal arising from the
difficulty of explaining outcomes that contradict prior mental
models~\cite{Samin2025SurpriseAdapt}.
It acts as a detector of epistemic novelty, alerting the agent to
incomplete knowledge without assessing severity of harm.
\emph{Cognitive regret}, by contrast, provides the evaluative
dimension: a counterfactual comparison between the chosen action
and a superior alternative~\cite{Buchanan2016TheRegret,
Zeelenberg1996ConsequencesMaking}.

\textbf{How surprise is operationalized.}
In \textsc{ms-rgr}, surprise is a binary flag
$\textit{Surprise}_E \in \{0,1\}$ computed by the runtime monitor:
$\textit{Surprise}_E = 1$ when a runtime observation deviates from
any anticipated outcome branch in the \textsc{kaos} goal model by a
domain-calibrated margin~$\delta$ specified at design time as part
of the obstacle definition~\cite{VanLamsweerde2000HandlingEngineering}.
The margin~$\delta$ is set during Step~2 workshops
(Section~\ref{sec:step2}) by mapping observable system signals to
specific goal--obstacle pairs.
When repeated surprise signals occur, Step~6
(Section~\ref{sec:step6}) triggers \textsc{kaos} elaboration of
the obstacle hierarchy.

The key novelty lies in their \emph{intersection}: relying solely
on surprise risks unnecessary interruptions (the ``freezing robot''
problem)\cite{Amodei2016ConcreteSafety}, while regret alone is reactive and post-hoc.
\textsc{ms-rgr} intervenes only when both signals meet critical
thresholds, distinguishing benign anomalies (high surprise, low
regret: monitor without halting) from genuine hazards (high
surprise, high regret: escalate immediately).

\textbf{Gap in Early-Stage Requirements Engineering.}
Existing \textsc{re} frameworks manage known obstacles and generate
alternative resolutions~\cite{VanLamsweerde2000HandlingEngineering},
but lack mechanisms for systematically evaluating decision quality
under epistemic uncertainty\cite{Feng2024AnalyzingChecking}.
To our knowledge, no prior work has operationalized cognitive regret
as a systematic, early-stage \textsc{re} constraint that governs
autonomous, unsupervised interactions.

Existing \textsc{re4ai} research confirms that trust,
explainability, and evaluative safety remain underspecified for
autonomous systems~\cite{Habibullah2021Non-functionalIndustry,
Ahmad2023RequirementsStudy}.
Agentic-\textsc{ai} systems operate as ecosystems with emergent
collective behaviour that no individual agent's obstacle model
anticipates~\cite{Sapkota2026AIChallenges}, creating the specific
deployment-readiness gap \textsc{ms-rgr} fills without requiring
training data~\cite{Dalpiaz2020RequirementsIntelligence}.

\section{Related Work and Problem Framing}
\label{sec:related}
\textbf{A. Uncertainty in GORE and RE4AI.}
Recent systematic mapping studies highlight an unresolved
``\textsc{re}-deployment gap''~\cite{Habiba2024HowDirections,
Ahmad2023RequirementsStudy}.
Horkoff and Yu enable trade-off reasoning but lack continuous
runtime enforcement~\cite{Horkoff2016InteractiveEngineering}.
Cailliau and van Lamsweerde track probabilistic obstacle
satisfaction but rely strictly on environmental sensors, leaving
systems vulnerable to silent
failures~\cite{Cailliau2017RuntimeGoals}.

\textbf{B. Simulation Support for Goal Modelling.}
Heaven and Letier~\cite{Heaven2011SimulatingModels} show that
stochastic simulation of quantitative \textsc{kaos} goal models
enables design-time what-if analysis of alternative system designs;
\textsc{ms-rgr} instantiates this framework for evaluative safety
with the \textsc{dri} as quality variable.
Alrajeh~et~al.~\cite{Alrajeh2026Data-DependentSystems} identify
conditional monitorability~(R11) and probabilistic responsibility
assignment~(R13) as systematic \textsc{kaos} gaps for
\textsc{ml}-enabled systems; \textsc{ms-rgr} fills both.

\noindent\textbf{Positioning Against Related Work.}
The approaches most closely related to \textsc{ms-rgr}
share a common gap: the absence of a design-time
evaluative criterion for comparing resolution tactics
under epistemic uncertainty.
Van Lamsweerde and
Letier~\cite{VanLamsweerde2000HandlingEngineering}
provide \textsc{kaos} obstacle analysis but lack
explicit support for modelling and evaluating the
autonomy/reasoning trilemma problem; \textsc{ms-rgr} adds
the \textsc{dri} as a computable evaluative
attribute with the routine/reflective boundary
formally specified at design time.
Cailliau and van
Lamsweerde~\cite{Cailliau2017RuntimeGoals} provide
probabilistic runtime monitoring via \textsc{mape-k}
but require distributional data not always available
at deployment time and cannot detect silent
failure; \textsc{ms-rgr} computes \textsc{dri} from
day-1 interaction logs without prior distributions.
Horkoff et al.~\cite{Horkoff2019Goal-orientedStudy}
identify evaluative safety modelling as an open
direction without providing a mechanism; Dalpiaz and
Niu~\cite{Dalpiaz2020RequirementsIntelligence} and
Alrajeh et al.~\cite{Alrajeh2026Data-DependentSystems}
identify \textsc{re}-for-\textsc{ai} gaps including
conditional monitorability~(R11) and probabilistic
responsibility assignment~(R13) without resolving
them; \textsc{ms-rgr} contributes a concrete mechanism
that addresses R11 through the \textsc{dri}'s
log-based monitorability and R13 through the
Regret-Dominance Gate's deterministic tactic
assignment traceable to the goal--obstacle model.
Chechik~\cite{Chechik2019UncertainLearning} addresses
architectural \textsc{ml} uncertainty without the
behavioral deviation dimension that arises under
epistemic uncertainty at runtime;
Heaven and Letier~\cite{Heaven2011SimulatingModels}
provide quantitative \textsc{kaos} simulation without
an evaluative quality variable, \textsc{ms-rgr}
addresses both dimensions, instantiating Heaven and
Letier's framework with \textsc{dri} as quality
variable and
\textsc{tbr}/\textsc{tti}/\textsc{far}/\textsc{ic}
as objective functions.
Park et al.~\cite{Park2025DoGames} study regret as
an analytical property of \textsc{llm} agent behavior
post-hoc, without operationalizing it as a design-time
safety constraint; \textsc{ms-rgr} specifies regret
thresholds at design time, requiring no training data.

\noindent\textbf{Supporting the autonomy/reasoning
trilemma.}
A specific gap across the most directly relevant
approaches is the lack of explicit, computable support
for the autonomy/reasoning trilemma, the design-time
question of when an agentic system should act
autonomously versus switching to reflective reasoning or
human control\cite{Mitchell2025FullyDeveloped}.
Van Lamsweerde and
Letier~\cite{VanLamsweerde2000HandlingEngineering}
define resolution tactics for obstacles but provide
no criterion for deciding when autonomous execution
should yield to human oversight; the choice remains
implicit in the obstacle model.
Cailliau and van
Lamsweerde~\cite{Cailliau2017RuntimeGoals} partially
address this through sensor-driven adaptation in
\textsc{mape-k} but the boundary is probabilistic
and sensor-dependent, providing no safety guarantee
under silent sensor failure\cite{DeSanctis2026ARuntime}.
Samin et al.~\cite{Samin2025SurpriseAdapt} partially
address the trilemma via surprise-based triggers,
detecting epistemic novelty as a signal to switch
modes, but produce no safety signal under sensor
freeze, leaving the boundary unprotected when
$S_E{=}0$.
Park et al.~\cite{Park2025DoGames} provide post-hoc
regret analysis of agent decisions but do not
formalize the autonomy boundary as a pre-deployment
engineering constraint.
\textsc{ms-rgr} addresses this directly:
the Regret-Dominance Gate specifies, at design time,System~1 (routine autonomy, $\text{DRI} < \tau$),
System~2 (reflective reasoning, $\text{DRI} \geq \tau$)
or human escalation ($\text{DRI} \geq \tau_{\text{crit}}$), with the boundary
computable from interaction logs independently of
sensor state. Feasibility evidence for this boundary
under sensor freeze is provided in
Section~\ref{sec:eval} (RQ2, RQ5).

\section{A Multi-Signal Framework for Regret-Aware Goal Resolution
in Agentic-AI (MS-RGR)}

The method treats cognitive regret as a requirements-level indicator
of decision risk and uses it to guide the selection of safer
resolution tactics~\cite{Moerland2018EmotionSurvey}, particularly
in high-stakes domains such as elderly
care~\cite{Buchanan2016TheRegret,Zeelenberg1996ConsequencesMaking}.

\subsection{Overview}

The method is grounded in \textsc{gore}.
We adopt design-time obstacle analysis and resolution strategies
defined by van Lamsweerde and
Letier~\cite{VanLamsweerde2000HandlingEngineering}. We extend GORE modelling with our mechanism to provide requirements engineers with an approach to systematically evaluate these models for safety, where we use cognitive regret thresholds that constrain the admissible
resolution tactics for the obstacles.

The method consists of five main steps:
\begin{enumerate}
  \item Model goals and resolution tactics
  \item Identify regret-prone situations
  \item Define regret thresholds as safety requirements
  \item Trigger regret assessment via runtime signals
  \item Select regret-aware resolution tactics
\end{enumerate}

\noindent\textbf{Design-Time Guidance.}
\textsc{ms-rgr} answers three questions a requirements
engineer specifies before deployment:
\begin{itemize}
  \item \textit{When is autopilot/autonomy safe?}
        $\text{DRI}\!<\!\tau$ and $S_E\!=\!0$
        $\Rightarrow$ System~1 (RT0--RT1, routine execution).
  \item \textit{When must the agent reason reflectively?}
        $\text{DRI}\!\geq\!\tau$
        $\Rightarrow$ System~2 (RT2--RT6, regardless of $S_E$).
  \item \textit{When must it stop operation or escalate to a human?}
        $\text{DRI}\!\geq\!\tau_{\text{crit}}$
        $\Rightarrow$ RT7 (emergency escalation, human override).
\end{itemize}
All three predicates are computable from interaction
logs before deployment; none requires sensor
confirmation. RQ2 (Section~\ref{sec:eval}) tests
the critical case: the third predicate fires
correctly under sensor freeze
($S_E\!=\!0$), with TBR\,=\,0\% across all
100 seeds.

\begin{figure}[t]
  \centering
  \includegraphics[width=\columnwidth]{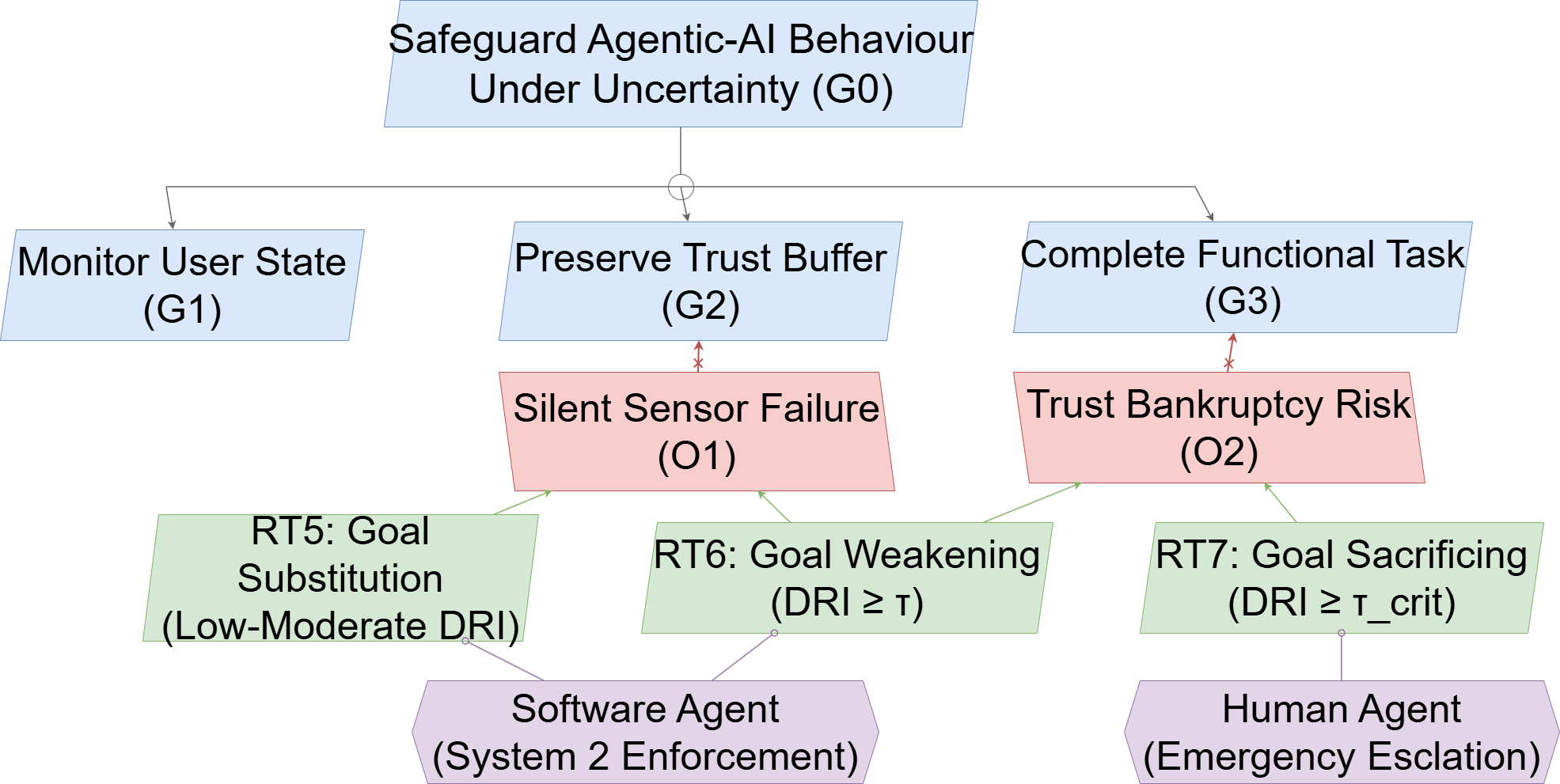}
  \caption{Design-time \textsc{kaos} goal-obstacle model
    establishing formal safety boundaries for \textsc{ms-rgr}.
    End-to-end traceability from G0 to obstacle-resolution tactics
    (\textsc{rt}5--\textsc{rt}7); latent risks O1 (silent sensor
    failure) and O2 (trust bankruptcy) are formally captured.}
  \label{fig:kaos_model}
\end{figure}

Figure~\ref{fig:kaos_model} presents the \textsc{ms-rgr}
\textsc{kaos} goal model~\cite{VanLamsweerde2000HandlingEngineering}.
Root goal~G0 is AND-refined into G1 (monitor user state),
G2 (preserve trust buffer), and G3 (complete functional task).
Obstacles O1 (silent sensor failure) and O2 (trust bankruptcy risk)
are resolved via \textsc{rt}5--\textsc{rt}7 under
regret-dominance.
The blue dashed path operationalises the \textsc{mape-k} feedback
loop~\cite{Cailliau2017RuntimeGoals}.

\begin{figure}[t]
  \centering
  \includegraphics[width=\columnwidth]{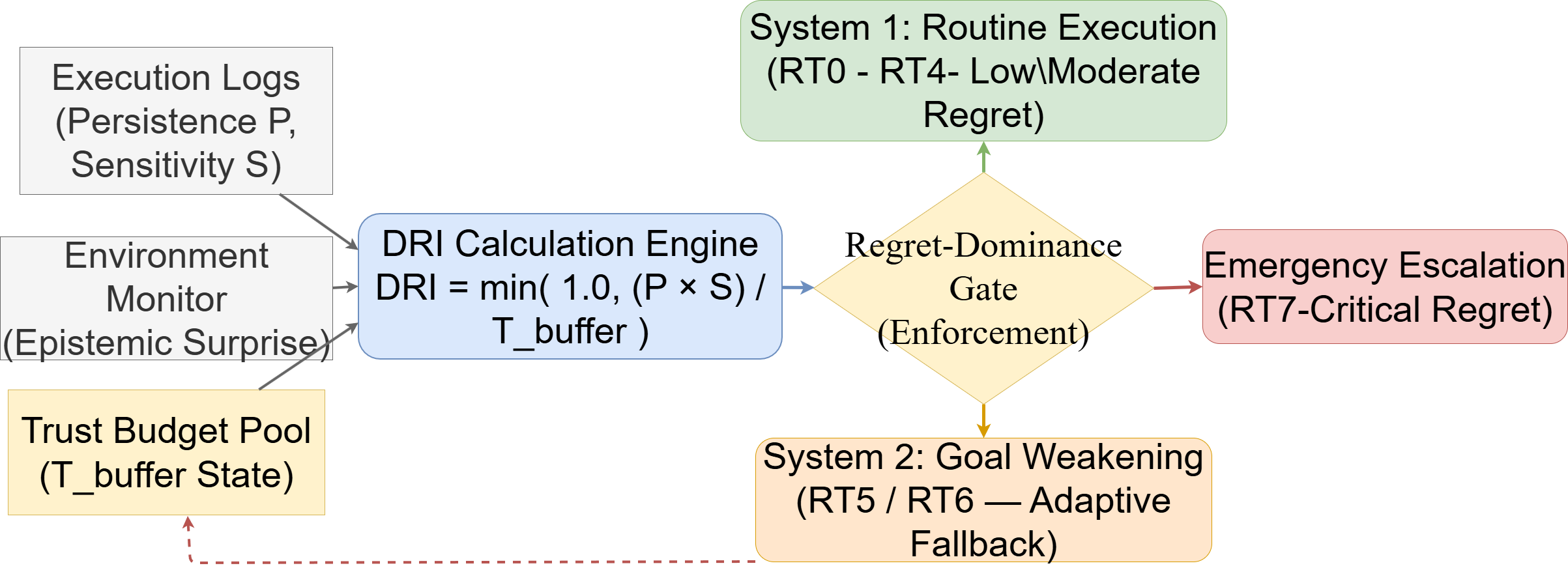}
  \caption{The \textsc{ms-rgr} runtime execution mechanism operationalizing the design-time requirements into an executable
    \textsc{mape-k}
    loop~\cite{Cailliau2017RuntimeGoals}.
    The \textsc{dri} synthesizes environmental surprise and execution
    logs, feeding into the Regret-Dominance Gate to route the agent across System~1 (routine), System~2 (reflective), and 
human escalation.}
  \label{fig:runtime_flow}
\end{figure}

\subsection{Step 1: Model Goals and Resolution Tactics}

System goals are captured using
\textsc{gore}/\textsc{kaos}~\cite{VanLamsweerde2000HandlingEngineering,
VanLamsweerdeReasoning}.
For each goal, potential obstacles are identified and multiple
resolution tactics are defined.
This follows established obstacle refinement practices in
\textsc{gore}~\cite{VanLamsweerdeGoal-OrientedTour,
VanLamsweerde2000HandlingEngineering}.

\subsection{Step 2: Identify Regret-Prone Situations}
\label{sec:step2}

For each resolution tactic, designers analyze situations where a
better alternative could have been selected, operationalizing
cognitive regret~\cite{Zeelenberg1996ConsequencesMaking}.
This is achieved through structured workshops using an anti-goal
and obstacle-based perspective: \textit{``How could the system
succeed functionally but fail ethically or
safely?''}~\cite{VanLamsweerdeReasoning}

\textbf{Scenario Template:}
\begin{itemize}
  \item \textbf{Context}: environmental or operational conditions
  \item \textbf{Action$_{\text{chosen}}$}: system action executed
  \item \textbf{Outcome$_{\text{actual}}$}: resulting outcome
  \item \textbf{Action$_{\text{alternative}}$}: feasible alternative
  \item \textbf{Outcome$_{\text{better}}$}: outcome if alternative
    had been chosen
\end{itemize}

\subsection{Step 3: Define Regret Thresholds as Safety Requirements}\label{sec:step3}

\textbf{Persistence~($P$)} is derived from accumulated count~$n$
of agent-initiated prompts within a sliding window of $W$ timesteps:
\begin{equation}
  P(n) \;=\; P_0 \cdot e^{\,\lambda n}
  \label{eq:persistence}
\end{equation}
where $P_0 = 0.1$ (default baseline) and $\lambda$ is the
reactance sensitivity coefficient (default~$0.15$; life-critical
domains use $\lambda = 0.25$).
The exponential model reflects how repeated unwanted interruptions
compound non-linearly, consistent with Brehm's Psychological
Reactance Theory~\cite{Brehm1989PsychologicalApplications}.

\textbf{User Sensitivity~($S$)} is a weighted composite:
\begin{equation}
  S \;=\; \alpha \cdot L \;+\; \beta \cdot R \;+\; \gamma \cdot A
  \label{eq:sensitivity}
\end{equation}
where $L$ = normalised response latency ratio, $R$ = refusal rate,
$A$ = abandonment rate; defaults $\alpha\!=\!0.5$, $\beta\!=\!0.3$,
$\gamma\!=\!0.2$, proportional to
\textsc{utaut}'s factor
rankings~\cite{Venkatesh2003UserVIEW1}.
All three signals are computable from standard deployment logs.

\textbf{Trust Buffer~($T_{\text{buffer}}$)} is initialised at~$1.0$
and updated after each agent action:
\begin{equation}
  T_{\text{buffer}}(t{+}1)
  \;=\;
  T_{\text{buffer}}(t) \times \bigl(1 - \eta \cdot \mathrm{DRI}_{t}\bigr)
  \label{eq:tbuffer}
\end{equation}
where $\eta = 0.05$ (conservative decay default;
sensitivity reported in supplementary material).
$T_{\text{buffer}}$ is bounded below at $T_{\min} = 0.05$,
encoding the near-ruin loss aversion property of Prospect
Theory~\cite{Kahneman1979ProspectRisk}:
$\mathrm{DRI} \to 1.0$ as $T_{\text{buffer}} \to T_{\min}$.

The \textsc{dri} synthesizes these signals:
\begin{equation}
  \mathrm{DRI}
  \;=\; \min\!\left(1.0,\;
        \frac{P(n) \times S}{T_{\text{buffer}}}\right)
  \label{eq:dri}
\end{equation}

\begin{example}[Elderly care medication scenario]
A resident takes $15$~s to acknowledge a prompt (expected: $5$~s),
giving $L = 3.0$, $R = 0.2$, $A = 0.1$:
$S = 0.5(3.0)+0.3(0.2)+0.2(0.1) = 1.58$.
After two unanswered prompts ($n=2$, $T_{\text{buffer}}=0.8$):
$P(2) = 0.1 \cdot e^{0.3} \approx 0.135$;
$\mathrm{DRI} \approx 0.27 < \tau = 0.6$ (System~1 continues).
After five prompts ($n=5$, $T_{\text{buffer}}=0.6$):
$P(5) \approx 0.212$;
$\mathrm{DRI} \approx 0.56$ (Medium zone, gate primes System~2).
\end{example}

\textbf{Threshold derivation.}
Three ordinal zones emerge from the $P \times S / T_{\text{buffer}}$
structure: \emph{Low} ($<0.4$): routine execution safe;
\emph{Medium} ($0.4 \leq \text{DRI} < 0.6$): adaptive tactics
appropriate; \emph{High} ($\geq 0.6$): by Prospect
Theory~\cite{Kahneman1979ProspectRisk} the expected loss outweighs
the gain; safety-enforcing tactics mandated.
$\tau\!=\!0.6$ is the High-zone boundary derivable
from the $P\!\times\!S/T_{\text{buffer}}$ structure
(Section~\ref{sec:step3}); domain variants
$\tau\!=\!0.4$ (life-critical) and $\tau\!=\!0.8$
(low-stakes) bracket this default, with $\tau\!=\!0.8$
empirically identified as unsafe per  Section~\ref{sec:eval}).

\subsection{Step 4: Detect Epistemic Novelty and Initiate
Multi-Signal Reflection}

Runtime monitoring detects epistemic novelty as a trigger for
reflective reassessment of goals and obstacles.
Decision authority remains anchored in the design-time regret
thresholds; even without surprise, safety is preserved via
\textsc{dri} accumulation.

\subsection{Step 5: Regret-Aware Selection of Resolution Tactics}\label{sec:step5}

The gating function $G(t)$ is evaluated at each timestep through
three deterministic steps.

\textbf{Step~G1 --- Compute \textsc{dri}.}
From log values $L$, $R$, $A$, $n$, $T_{\text{buffer}}(t)$,
compute $S$ via Eq.~\eqref{eq:sensitivity},
$P(n)$ via Eq.~\eqref{eq:persistence},
and \textsc{dri} via Eq.~\eqref{eq:dri}.

\textbf{Step~G2 --- Route to admissible tactic class.}
\begin{align}
  \lefteqn{\textbf{[Regret Dominance---Critical]}} \notag \\
  &\quad \text{DRI}(t) \geq \tau_{\text{crit}}
    \;\Rightarrow\; \textsc{rt} \in \{\textsc{rt}7\} \\[3pt]
  \lefteqn{\textbf{[System~2---Protective]}} \notag \\
  &\quad \tau \leq \text{DRI}(t) < \tau_{\text{crit}}
    \;\Rightarrow\; \textsc{rt} \in \{\textsc{rt}5,\textsc{rt}6\} \\[3pt]
  \lefteqn{\textbf{[System~2---Adaptive]}} \notag \\
  &\quad \textit{Surprise}_E{=}1 \wedge \text{DRI}(t) < \tau
    \;\Rightarrow\; \textsc{rt} \in \{\textsc{rt}2\text{--}4\} \\[3pt]
  \lefteqn{\textbf{[System~1---Routine]}} \notag \\
  &\quad \text{DRI}(t) < \tau \wedge \textit{Surprise}_E{=}0
    \;\Rightarrow\; \textsc{rt} \in \{\textsc{rt}0,\textsc{rt}1\}
\end{align}

\textbf{Step~G3 --- Update trust buffer} via Eq.~\eqref{eq:tbuffer}.

Regret dominates over surprise because evaluative risk is a
design-time guarantee that cannot be silenced by sensor failure.
When $\textit{Surprise}_E = 0$, the gate fires at Step~G2 when
$\text{DRI}(t) \geq \tau_{\text{crit}}$ because $P$ accumulates
from interaction logs independently of the frozen sensor.Every tactic selection therefore, traces to the design-time
goal--obstacle model~\cite{Cailliau2017RuntimeGoals,
Habibullah2021Non-functionalIndustry}, making the
routine/reflective switch formally auditable before deployment.

\textbf{Formal Safety Properties.}
Table~\ref{tab:ltl_props} lists the five 
global \textsc{ltl} safety properties.
\begin{table}[H]
\caption{\textsc{ms-rgr} Global \textsc{ltl} Safety Properties}
\label{tab:ltl_props}
\resizebox{\columnwidth}{!}{
\centering\small
\begin{tabular}{lp{3.2cm}p{3.0cm}}
\toprule
Property & \textsc{ltl} Formula & Interpretation \\
\midrule
Regret Dominance
  & $\Box(\text{DRI} \geq \tau_{\text{crit}} \to \bigcirc\textsc{rt}7)$
  & Critical regret mandates \textsc{rt}7 \\
System~2 Trigger
  & $\Box(\text{DRI} \geq \tau \to \Diamond S_2)$
  & High regret triggers reflective mode \\
Trust Invariance
  & $\Box(\text{DRI} \le 1.0)$
  & Regret bounded below total depletion \\
Silent Failure
  & $\Box(\neg S_E \wedge \text{DRI} \geq \tau \to \Diamond\textsc{rt}6\text{-}7)$
  & Regret overrides silent sensor \\
Surprise Response
  & $\Box(S_E > 0 \to \bigcirc(S_2 \vee \text{DRI\_Up}))$
  & Anomalies trigger analysis or risk update \\
\bottomrule
\end{tabular}
}
\end{table}

\textbf{Residual Risk Labels.}
Once $t_{\text{optimal}}$ is chosen, it receives a residual risk
label~\cite{2013SoftwareII,Westhofen2023CriticalityArt}:
M0~(Normal), M1~(Cautious), M2~(Interrupt).

\subsection{Step 6: Feedback for Model Evolution}
\label{sec:step6}

\begin{figure}[t]
  \centering
  \includegraphics[width=\columnwidth]{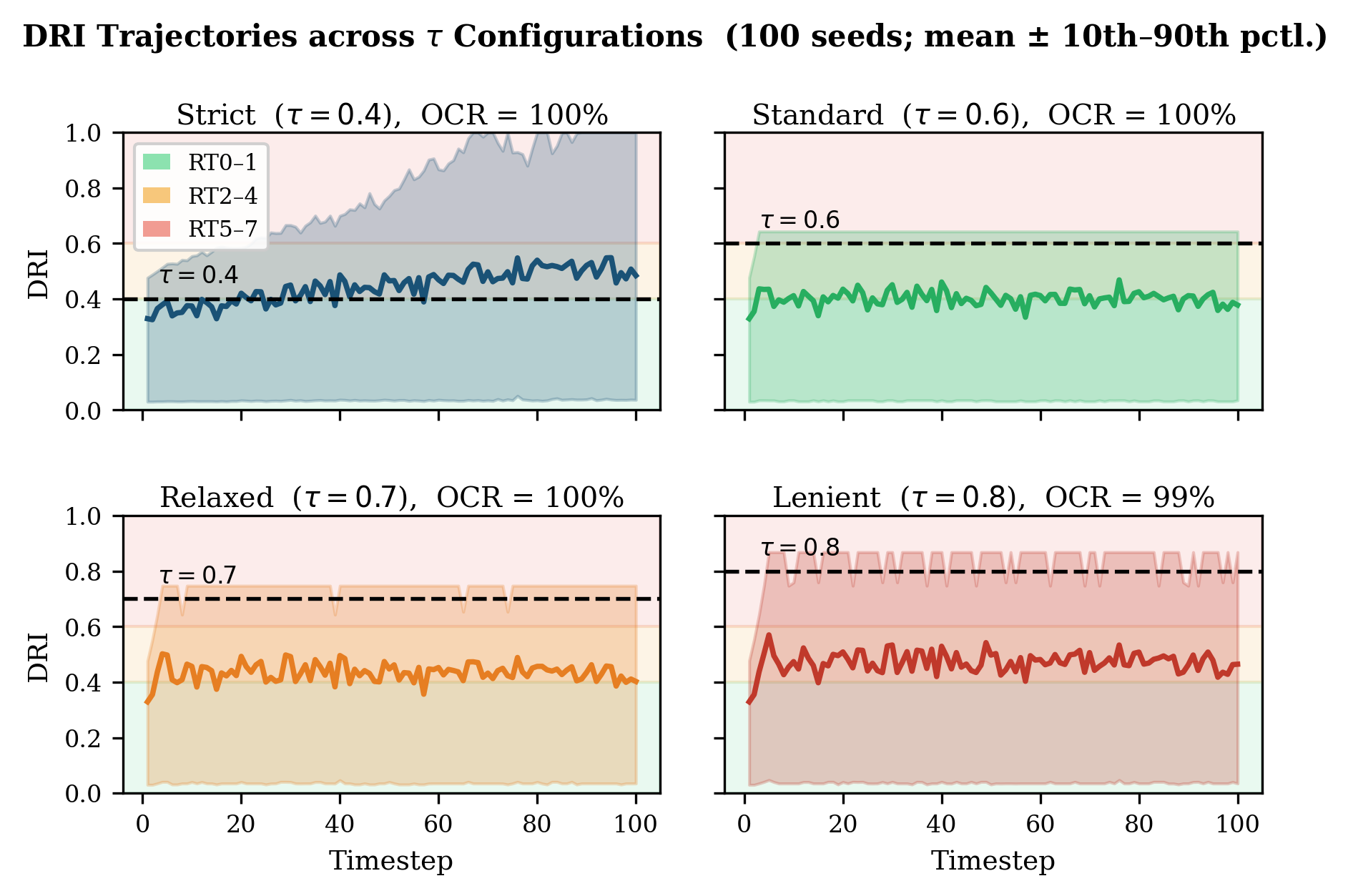}
  \caption{DRI values over 100 timesteps across four $\tau$ configurations
    (100 seeds; mean $\pm$ 10th--90th percentile band).
    Three configurations maintain correct zone assignment across
    all seeds; Lenient mode ($\tau\!=\!0.8$) shows systematic
    DRI intrusion into the RT5--7 zone, directly explaining the
    OCR\,=\,99\% failure and the calibration warning (RQ1).
    Parameter sensitivity ($\lambda \in \{0.10\text{--}0.25\}$,
    $\eta \in \{0.03\text{--}0.10\}$) confirms ordinal zone
    ordering is preserved across the full parameter range
    (\url{https://doi.org/10.5281/zenodo.20928801}).}
  \label{fig:dri_trace}
\end{figure}

Repeated or persistent surprise signals trigger \textsc{kaos}
obstacle elaboration rather than runtime decision
inputs~\cite{VanLamsweerde2000HandlingEngineering,
Letier2025ObstacleChallenges}.
Engineers return to the goal--obstacle model to elaborate the newly
observed condition as a refined obstacle, applying goal elaboration,
obstacle decomposition, or tactic substitution.Figure~\ref{fig:dri_trace} shows the resulting 
DRI accumulation over 100 timesteps across four 
$\tau$ configurations.

\subsection{Regret-Calibrated Resolution Tactics}
Table~\ref{tab:regret_tactics} summarizes the regret resolution tactics.
\begin{table}[H]
\caption{Regret Levels and Resolution Tactics (\textsc{rt}0--\textsc{rt}7).
  \textsc{rt}5--\textsc{rt}7 correspond in \textsc{gore} terms to
  Goal Substitution~\cite{VanLamsweerdeReasoning},
  Goal Weakening~\cite{Letier2004ReasoningEngineering},
  and Goal Sacrificing~\cite{VanLamsweerde1998ManagingEngineering}.}
\label{tab:regret_tactics}
\centering\small
\begin{tabular}{p{1.8cm}p{1.0cm}p{4.5cm}}
\toprule
Regret Level & Tactic & Description \\
\midrule
Low
  & RT0 & Continue execution with passive monitoring \\
  & RT1 & Log anomaly; no intervention \\
\midrule
Medium
  & RT2 & Conservative adaptation (slower/safer mode) \\
  & RT3 & Request additional sensing or explanation \\
  & RT4 & Notify human or peer agent (non-blocking) \\
\midrule
High
  & RT5 & Interrupt current execution \\
  & RT6 & Switch to fail-safe configuration \\
  & RT7 & Escalate to human or emergency protocol \\
\bottomrule
\end{tabular}
\end{table}

\subsection{Handling Undetected Surprise Signals}

When $S_E{=}0$ despite genuine risk, high \textsc{dri} triggers
M1-Cautious (escalate to intermediate \textsc{rt} classes);
low \textsc{dri} maintains M0-Normal while logging for future
elaboration (Step~\ref{sec:step6}).
This ensures conservative behaviour even when epistemic monitoring
fails --- the S3 scenario confirmed in
Section~\ref{sec:eval}.

\section{Method Instantiation: Elderly Care}

This instantiation applies \textsc{ms-rgr} 
to an elderly-care system across mobility 
assistance, medication adherence, and 
emergency response.

\subsection{Goals and Obstacles}

System goals follow \textsc{kaos} principles:
G1 (safe mobility), G2 (medication compliance),
and G3 (timely emergency response), each with identified obstacles
representing regret-prone situations.

Table~\ref{tab:rt_elderly} instantiates 
the tactics for the elderly care domain.
\begin{table}[h]
\centering\small
\caption{\textsc{rt}0--\textsc{rt}7 Tactic Classes: Elderly Care}
\label{tab:rt_elderly}
\begin{tabular}{cp{5.5cm}}
\toprule
\textbf{RT} & \textbf{Illustrative Tactic Class} \\
\midrule
RT0 & Passive monitoring and internal logging \\
RT1 & Non-intrusive verbal prompt to resident \\
RT2 & Context-aware reminder with adaptive timing \\
RT3 & Escalated reminder requiring acknowledgment \\
RT4 & Notification to human caregiver \\
RT5 & Automated safety check or precautionary intervention \\
RT6 & Explicit caregiver intervention request \\
RT7 & Emergency escalation (e.g., emergency services) \\
\bottomrule
\end{tabular}
\end{table}

\subsection{Regret-Prone Situations}

\textbf{Representative Example --- Mobility Assistance:}
A verbal prompt (RT1) is selected while a resident attempts to
stand, resulting in a fall risk.
This outcome is evaluated as High regret, indicating that RT4
(caregiver notification) would have been preferable.
The DRI numerical trace confirming this assignment is given in
Example~1 (Section~\ref{sec:step2}).

\subsection{Runtime Obstacle Handling}

Runtime tactic selection uses $G(t)$: high \textsc{dri} with or
without $S_E$ escalates to \textsc{rt}3--\textsc{rt}7 (M1 or~M2);
low \textsc{dri} maintains \textsc{rt}0--\textsc{rt}1 (M0)
regardless of surprise state.

\section{Method Instantiation: Autonomous Driving}

To illustrate domain-independence, we instantiate
\textsc{ms-rgr} in a passenger comfort scenario.
When \textsc{dri} reaches $0.64 \geq \tau = 0.6$
during fast cornering, the gate routes to RT6
(Goal Weakening~\cite{Letier2004ReasoningEngineering}),
preserving partial goal satisfaction via the same
gate logic and RT0--RT7 structure as the elderly-care
domain, with \textsc{dri} variables re-instantiated
over navigation signals rather than care-interaction
logs. The full \textsc{mape-k} trace and \textsc{dri}
re-instantiation are provided in the repository, supporting the domain-agnostic
structure of Equations~(1)--(4).
\section{Evaluation}
\label{sec:eval}
We conduct a feasibility study of \textsc{MS-RGR} using goal-driven stochastic simulation following
Design Science Research methodology~\cite{Hevner2004DESIGN1}, consistent with
analogous RE feasibility studies~\cite{Chu2024IntegratingAdaptation}.
Building on Heaven and Letier's framework for simulating quantitative goal
models~\cite{Heaven2011SimulatingModels}, the DRI serves as the quality variable
computed from standard execution logs $(L, R, A, n)$.
The simulator allows requirements engineers to conduct proactive what-if
analyses of safety thresholds~$(\tau)$ and resolution tactics prior to
deployment.

\subsection{Experiment Setup}

\noindent\textbf{Simulation parameters.}
We ran the simulation across \textbf{100 independent random seeds} over
\textbf{100 discrete timesteps} per seed.
At each timestep two probabilistic events are injected independently:
\begin{itemize}
  \item \textbf{Friction} at $\Pr(\text{friction}) = 0.7$, reflecting
        documented user-resistance rates~\cite{Brehm1989PsychologicalApplications}.
  \item \textbf{Surprise} at $\Pr(\text{surprise}) = 0.2$, reflecting
        silent-failure rates in LLM agent benchmarks~\cite{Andriushchenko2025AgentHarm:Agents,Zhang2025AGENT-SAFETYBENCH:Agents}.
\end{itemize}
The DRI is computed each step via Equations~(1)--(4), and the
Regret-Dominance Gate $G(t)$ selects a resolution tactic from RT0--RT7.
Seven metrics structure the evaluation (Table~\ref{tab:metrics}).

\begin{table}[htbp]
\centering
\caption{Evaluation metrics and the research question each addresses.}
\label{tab:metrics}
\resizebox{\columnwidth}{!}{
\begin{tabular}{lll}
\toprule
\textbf{Metric} & \textbf{RQ} & \textbf{Meaning} \\
\midrule
OCR (Ordinal Calibration Rate) & RQ1 & DRI produces correct RT tiers \\
TBR (Trust Bankruptcy Rate)    & RQ2 & Silent failure prevented \\
TTI (Time-to-Intervene)        & RQ2 & Speed of first detection \\
LPR (LTL Pass Rate)            & RQ3 & Decisions traceable to KAOS \\
ACT\% (Activation Rate)        & RQ4 & Gate escalation frequency \\
FAR\% (False Alarm Rate)       & RQ4 & Spurious escalations \\
IC (Intervention Cost)         & RQ4 & Mission disruption per step \\
ACT\%/FAR\%/IC (re-examined)   &RQ5  & Ablation-derived structural
                                      coverage evidence\\
\bottomrule
\end{tabular}
}
\end{table}


\noindent\textbf{Research questions.}
Five research questions structure the evaluation:
\begin{itemize}
  \item \textbf{RQ1 (Operability):} Can the DRI engine compute correctly
        ordered regret levels from standard logs without human subject data?
  \item \textbf{RQ2 (Robustness):} Does the Regret-Dominance Gate maintain
        safe tactic selection consistent with the specified safety thresholds when $S_E = 0$
        (the sensor-freeze S3 scenario)?
  \item \textbf{RQ3 (Traceability):} Are all runtime tactic selections
        formally traceable to the design-time KAOS model via LTL
        properties P1--P5?
  \item \textbf{RQ4 (What-If Utility):} Does stochastic simulation
        correctly identify the contribution of each behavioral theory to
        goal-refinement decisions?
  \item \textbf{RQ5 (Systematic Coverage):} To what extent does the
  Regret-Dominance Gate's ordinal zone structure provide
  systematic, graduated coverage of the routine/reflective/escalation trilemma, compared to an unstructured (single-threshold or
  binary) approach?
\end{itemize}

\noindent\textbf{What-If Analysis~I (RQ1--RQ3).}
Three structural conditions isolate each component:
\begin{itemize}
  \item \textbf{A -- Full \textsc{MS-RGR}:}
        DRI engine + Regret-Dominance Gate + dynamic $T_{\text{buf}}$.
  \item \textbf{B -- Fixed-Trust Baseline:}
        DRI with static $T_{\text{buf}}$ ($\eta\!=\!0$);
        no Prospect Theory loss-aversion update~\cite{Kahneman1979ProspectRisk}.
  \item \textbf{C -- Sensor-Only Baseline:}
        Surprise-threshold gating only; no DRI.
        Represents prior epistemic-only approaches~\cite{Cailliau2017RuntimeGoals,Samin2025SurpriseAdapt}.
\end{itemize}

\noindent\textbf{What-If Analysis~II (RQ4 -- Ablation).}
One behavioral theory is removed at a time
($\tau\!=\!0.6$, $\tau_{\text{crit}}\!=\!0.8$)~\cite{Alrajeh2026Data-DependentSystems}:
\begin{itemize}
  \item \textbf{B$'$ -- No Prospect Theory:}
        $T_{\text{buf}}$ fixed ($\eta\!=\!0$);
        no trust decay or recovery~\cite{Kahneman1979ProspectRisk}.
  \item \textbf{C$'$ -- No Reactance Theory:}
        Linear $P$ ($+0.2$ per friction event) instead of
        exponential~\cite{Brehm1989PsychologicalApplications}.
  \item \textbf{D -- No Dual-Process:}
        $\tau$ gate removed; always escalate to System~2.
  \item \textbf{E -- No Satisficing:}
        RT6 replaced by RT7; goal-weakening step
        skipped~\cite{Letier2004ReasoningEngineering,Simon1955REPRINTED:Man}.
\end{itemize}

\subsection{Case Study I: Goal-Driven Stochastic Simulation}

The first case study evaluates \textsc{MS-RGR} over the stochastic
simulation described above.
RQ1--RQ3 correspond to What-If Analysis~I; RQ4 to the ablation study.

\subsubsection{RQ1 -- Operability}

In all 100 seeds, DRI was computed from synthetic interaction logs
$(L, R, A, n)$ with no human subject input.
OCR = 100\% in Strict ($\tau\!=\!0.4$), Standard ($\tau\!=\!0.6$), and
Relaxed ($\tau\!=\!0.7$) configurations: RT0--1 fired exclusively at
DRI\,$<$\,0.4, RT2--4 at $0.4\!\leq\!$DRI\,$<$\,0.6, and RT5--7 at
DRI\,$\geq$\,0.6, with zero ordering inversions.
In Lenient mode ($\tau\!=\!0.8$), OCR = 99\%, confirming
$\tau \leq 0.7$ is required for safe deployments.

\smallskip
\textit{RQ1:The DRI engine reliably produces correctly
ordered regret levels from standard logs.}

\subsubsection{RQ2 -- Robustness}

Table~\ref{tab:whif1} reports TBR and TTI
under the three structural conditions.
Full \textsc{MS-RGR} yielded TBR\,=\,0\% versus 96\%
under the Sensor-Only baseline during S3 sensor-freeze scenarios
($p\!<\!0.001$, Wilcoxon signed-rank test~\cite{Durango2018AnTest},
100 seeds).
Even when $S_E = 0$ throughout, Condition~A maintained safety via the
Silent-Failure LTL property:
\[
  \square\!\left(\neg S_E \;\wedge\; \mathrm{DRI} \geq \tau
    \;\rightarrow\; \diamond\,\mathrm{RT}_{6\text{-}7}\right)
\]
Full \textsc{MS-RGR} also detected risk 17.5$\times$ faster than
Sensor-Only (4.1 vs.\ 72.0 timesteps).
The 34\% TBR of Condition~B confirms that dynamic
$T_{\text{buf}}$~\cite{Kahneman1979ProspectRisk} is structurally essential:
without Prospect Theory loss-aversion trust recovery, one in three
runs end in silent failure.
Effect sizes are large by conventional standards: Cohen's
$h\!=\!2.74$ for the A-vs-C TBR comparison (arcsine-transformed
proportions $0\%$ vs.\ $96\%$) and $h\!=\!1.24$ for A-vs-B
($0\%$ vs.\ $34\%$), both exceeding the $h\!=\!0.8$ large-effect
threshold~\cite{Cohen1988StatisticalEdition}.
For TTI, rank-biserial $r\!=\!1.00$ (all 100 seeds ordered
A\,$<$\,C consistently), confirming a practically decisive effect.
The 17.5$\times$ TTI speedup (4.1 vs.\ 72.0 steps) maps to
intervening at timestep~4 rather than timestep~72 across a 100-step
mission horizon, operationally meaningful in both elderly-care and
autonomous-driving deployments.

\begin{table}[htbp]
\centering\footnotesize
\setlength{\tabcolsep}{3.5pt}
\caption{What-If Analysis~I: TBR, TTI, effect size, and S3 safety
  outcome (100 seeds $\times$ 100 steps, $\tau\!=\!0.6$,
  $\Pr(\text{friction})\!=\!0.7$, $\Pr(\text{surprise})\!=\!0.2$;
  \texttt{whif1\_simulation.py}).}
\label{tab:whif1}
\begin{tabular}{@{}lcccc@{}}
\toprule
\textbf{Condition} & \textbf{TBR} & \textbf{TTI} & \textbf{Effect Size$^a$} & \textbf{S3} \\
\midrule
A: Full \textsc{MS-RGR} & 0\%  & 4.1  & ---                         & Safe \\
B: Fixed-Trust          & 34\% & 28.7 & $h\!=\!1.24$                & Partial \\
C: Sensor-Only          & 96\% & 72.0 & $h\!=\!2.74$, $r\!=\!1.00$ & \textbf{Unsafe} \\
\midrule
\multicolumn{5}{@{}l@{}}{\footnotesize
  $^{a}$vs.\ Condition~A; Cohen's $h$ on TBR~\cite{Cohen1988StatisticalEdition},
  $r$ rank-biserial on TTI.}\\
\bottomrule
\end{tabular}
\end{table}

\smallskip
\textit{RQ2:Regret-Dominance reduced silent-failure incidence
under the simulated conditions studied
($p\!<\!0.001$) and responds 17.5$\times$ faster than single-signal
epistemic approaches.}

\subsubsection{RQ3 -- Traceability}

\begin{figure}[t]
\centering
\includegraphics[width=\columnwidth]{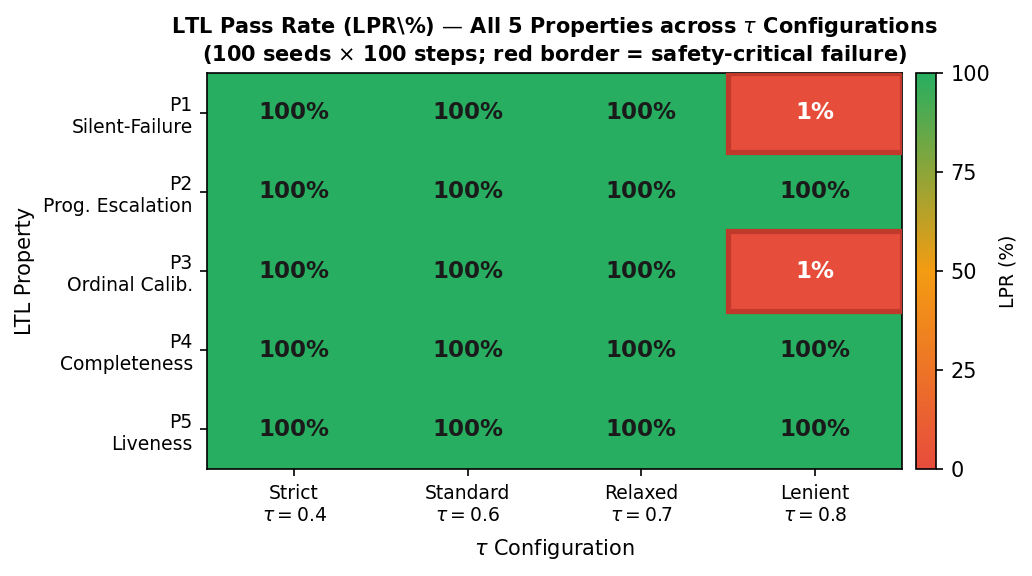}
\caption{%
  LTL Pass Rate for P1--P5 across all four $\tau$ configurations
  ($100 \times 4 = 400$ traces).
  Red-bordered cells indicate safety-critical failure
  (LPR\,$<$\,100\%).
  P1 (Silent-Failure) and P3 (Ordinal Calibration) fail at
  $\tau\!=\!0.8$, providing direct empirical evidence that
  $\tau\!\leq\!0.7$ is required for safe deployment (RQ3).}
\label{fig:ltl_matrix}
\end{figure}

All five LTL properties in figure  \ref{fig:ltl_matrix} achieved LPR = 100\% in Strict,
Standard, and Relaxed configurations across $100\!\times\!4 = 400$ traces,
confirming every runtime RT selection is formally traceable to the
design-time KAOS model via $G(t)$.
The near-total failure (LPR = 1\%) of P1 and P3 in Lenient mode
($\tau\!=\!0.8$) is an empirical calibration warning: $\tau\!=\!0.8$ is
unsafe for life-critical deployments.
This is precisely the kind of goal-model what-if analysis Heaven and
Letier's framework is designed to support~\cite{Heaven2011SimulatingModels}.

Notably, the Silent-Failure property
($\square(\neg S_E \wedge \text{DRI}\!\geq\!\tau
\to \diamond\text{RT}_{6\text{-}7})$,
Table~\ref{fig:ltl_matrix}) formally specifies the
autonomy/reasoning boundary under sensor freeze:
the System~1$\to$System~2 switch traces to the
\textsc{kaos} goal model independently of sensor
state, directly addressing the trilemma raised in
Section~\ref{sec:related}.

\smallskip
\textit{RQ3:All runtime adaptations are formally traceable
under properly calibrated configurations.}

\subsubsection{RQ4 -- What-If Utility}

To answer RQ4, we ask: \textit{what if the mechanism lacked one
behavioral theory?}
Figure~\ref{fig:ablation} reports results (full data in supplementary
Table~S1); each theory's non-redundant contribution is confirmed:
removing Prospect Theory raises ACT\% from 33.3\% to 52.5\%;
removing Reactance Theory delays TTI by 2.5 steps (42\% slowdown);
removing the dual-process gate raises FAR\% to 100\% ,every escalation
spurious; removing Satisficing raises IC by 30\%.
All four Wilcoxon tests return $p\!<\!0.001$.

\begin{figure}[htbp]
\centering
\includegraphics[width=\columnwidth]{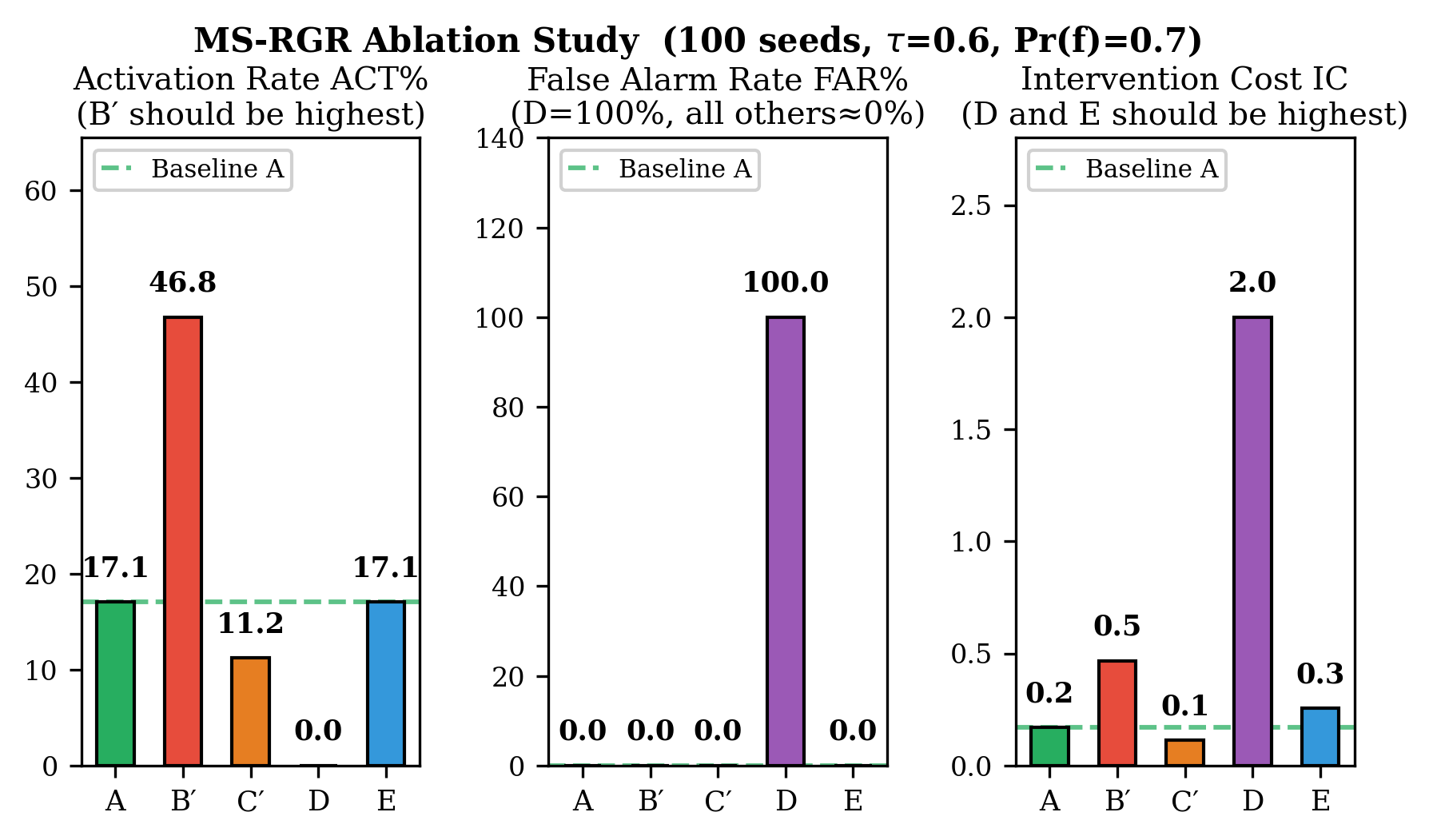}
\caption{%
  Theory ablation results (What-If Analysis~II).
  Each bar group contrasts Condition~A (Full \textsc{MS-RGR}) against
  the condition that removes one behavioral theory.
  Condition~A uniquely minimises all four metrics simultaneously
  ($p\!<\!0.001$, Wilcoxon tests).}
\label{fig:ablation}
\end{figure}

\noindent\textbf{B$'$ -- No Prospect Theory~\cite{Kahneman1979ProspectRisk}.}
Removing loss-aversion trust recovery (fixing $\eta\!=\!0$) raises
ACT\% from 33.3\% to 52.5\% ($p\!=\!0.001$ on ACT\%).
Without $T_{\text{buf}}$ restoration, trust depletions accumulate
across escalations, keeping DRI chronically elevated and triggering
the gate 1.6$\times$ more often than Condition~A.
\textit{Design implication}: specify dynamic $T_{\text{buf}}$ recovery
as a design-time KAOS requirement.

\smallskip
\noindent\textbf{C$'$ -- No Reactance Theory~\cite{Brehm1989PsychologicalApplications}.}
Replacing exponential persistence $P(n)$ with linear accumulation
raises mean TTI from 6.0 to 8.5 steps
($p\!=\!0.001$ on TTI), a 42\% slowdown in first detection.
Exponential $P(n)$ accelerates across friction events through
compounding growth; the linear model accumulates more slowly,
delaying threshold crossing for small~$n$.
\textit{Design implication}: specify exponential persistence; the
compounding model detects persistent threats 42\% sooner.

\smallskip
\noindent\textbf{D -- No Dual-Process.}
Removing the $\tau$ gate and always escalating to System~2 raises
FAR\% to 100\% and IC to 2.000 (4.2$\times$ higher than
Condition~A, $p\!=\!0.001$ on FAR\%).
Without the gate, forced escalations reset $P\!=\!P_0$ each step,
keeping DRI permanently below $\tau$ "the
\emph{freezing-robot}" problem~\cite{Amodei2016ConcreteSafety}: every
escalation is spurious yet no DRI threshold is ever crossed.
\textit{Design implication}: retain the $\tau$ threshold; unconditional
escalation is both wasteful and insensitive to genuine hazard
accumulation.

\smallskip
\noindent\textbf{E -- No Satisficing~\cite{Simon1955REPRINTED:Man}.}
Replacing RT6 with RT7 (skipping goal-weakening~\cite{Letier2004ReasoningEngineering})
raises IC by 30\% (0.626 vs.\ 0.480, $p\!=\!0.001$ on IC) with
identical ACT\% and FAR\%, confirming that the costlier resolution
tactic adds disruption without additional safety benefit.
\textit{Design implication}: specify RT6 before RT7; goal-weakening
reduces mission disruption 30\% at no safety cost, consistent with
bounded rationality~\cite{Simon1955REPRINTED:Man}.

\smallskip
\textit{RQ4:}
Condition~A uniquely achieves the lowest activation rate (33.3\%),
fastest joint detection (TTI\,=\,6.0 steps), zero false alarms, and lowest
intervention cost (IC\,=\,0.480) simultaneously
($p\!=\!0.001$ for all four Wilcoxon tests), providing feasibility
evidence that all four behavioral theories are necessary and
non-redundant components of \textsc{MS-RGR}.


\subsubsection{RQ5 -- RE Coverage of the routine/reflective/escalation trilemma}
\label{sec:re_coverage}
Systematic Coverage: \textsc{MS-RGR} addresses RQ5
through three structural properties derivable directly from
the mechanism design:
(i)~$\tau$ and $\tau_{\text{crit}}$ convert the informal
``when should the agent pause?'' question into a computable
predicate over observable logs, specified as KAOS design-time
constraints rather than runtime heuristics;
(ii)~the DRI range is partitioned into three ordinal zones
(Low, Medium, High; Table~III), each gated by a verified LTL
threshold property (Table~II), giving the gate a non-binary,
graduated response absent from single-threshold approaches; and
(iii)~every gate activation traces to a KAOS obstacle and LTL
property, making the trilemma boundary formally auditable at
design time.

The ablation study (Figure~\ref{fig:ablation}) offers indirect
evidence relevant to RQ5. Removing the dual-process gate
(Condition~D, FAR\,=\,100\%) collapses the mechanism into binary
behaviour with no intermediate zone: every interaction is routed
to RT7 regardless of actual DRI level. Condition~D demonstrates
that the ordinal zone structure, not merely the DRI computation
itself, is what enables differentiated, proportionate response:
a specification omitting this structure cannot represent the
transitional cases that RT2--RT4 are designed to handle.
Removing Prospect Theory (B$'$, ACT\%\,=\,52.5\%) produces
over-intervention -- the trust-recovery case is unspecified;
removing Reactance Theory (C$'$, TTI\,$+$2.5 steps) delays
detection of gradual escalation -- the compounding-persistence
case is unspecified; removing Satisficing (E, IC\,$+$30\%)
produces inconsistent resolution -- the proportionate-response
case is unspecified.

\begin{figure}[htbp]
\centering
\includegraphics[width=\columnwidth]{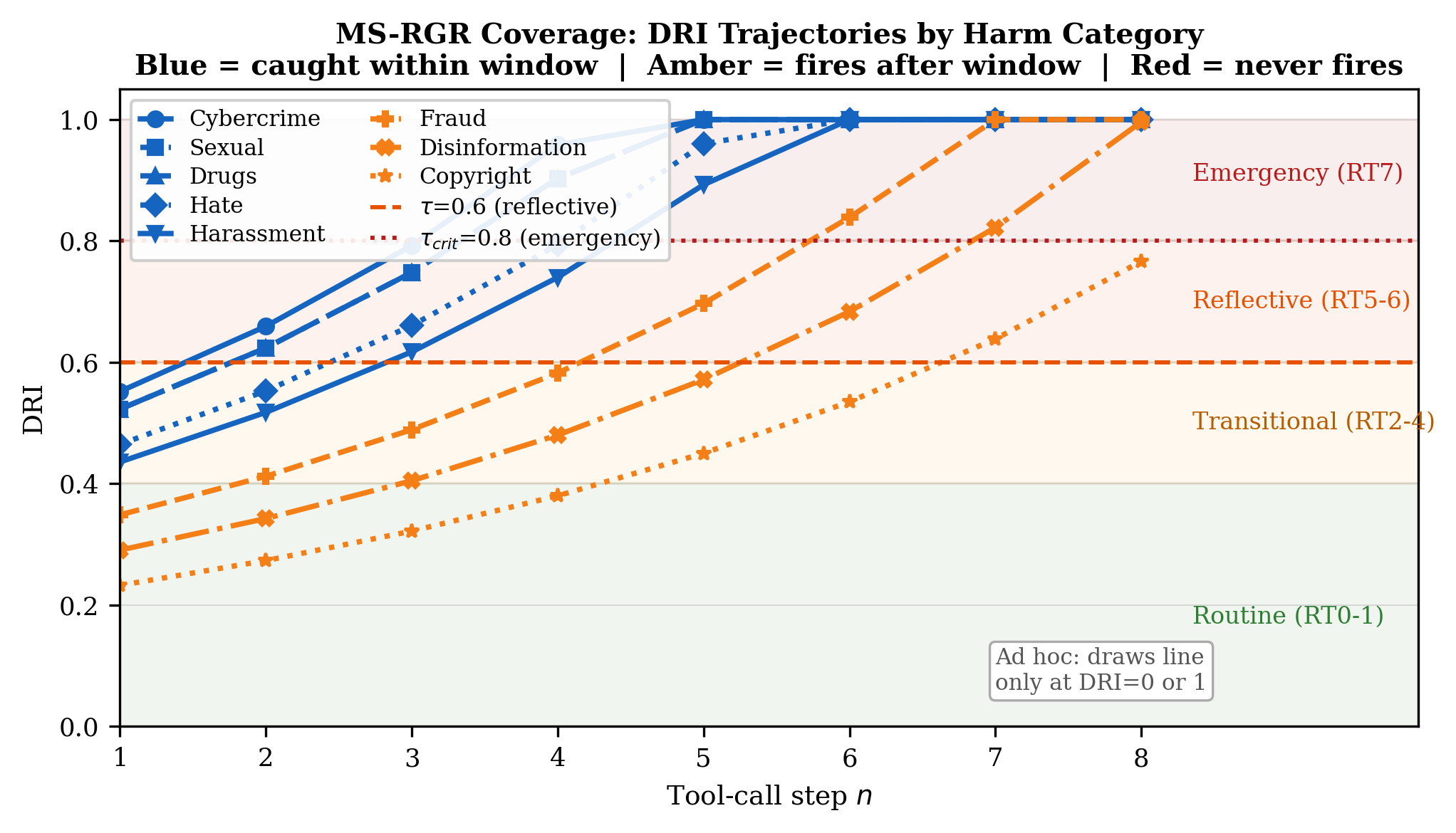}
\caption{Analytical DRI trajectories for eight AgentHarm harm
  categories over $n\!=\!1$--8 tool-call steps, derived from
  Equations~(1)--(4) at reported parameter defaults
  ($P_0\!=\!0.1$, $\lambda\!=\!0.15$, $\eta\!=\!0.05$,
  $\alpha\!=\!0.5$, $\beta\!=\!0.3$, $\gamma\!=\!0.2$).
  \textbf{Blue}: DRI crosses $\tau$ within the scenario window.
  \textbf{Amber}: DRI crosses $\tau$ after the scenario ends
  (calibration gap --- shorter $\tau$ or higher $L$ required).
  The figure is illustrative of mechanism behaviour, not a
  measurement of ad hoc practitioner coverage.
  Reproduction: \texttt{coverage\_analysis.py} in the
  replication package.}
\label{fig:coverage_dilemma}
\end{figure}

Figure~\ref{fig:coverage_dilemma} illustrates analytically,
from Equations~(1)--(4) at the reported parameter defaults,
how DRI accumulates differently across harm categories, showing
which categories enter the transitional zone within a typical
scenario window and which require longer interaction sequences
or lower $\tau$ to trigger intervention.

\textit{RQ5:}
Structural properties~(i--iii) and Condition~D's
binary collapse (FAR\,=\,100\%) jointly support
differentiated coverage of the routine/reflective/escalation trilemma. One limitation: FAR\% measures spurious
escalations relative to the \textsc{dri} threshold,
not ground-truth safety, activations where
DRI\,$\geq\!\tau$ but execution proves safe under
reflection constitute an unquantified second
false-alarm category. Medium-zone tactics (RT2--RT4)
partially mitigate this by responding to elevated
\textsc{dri} without breaking autonomy; full
characterisation requires ground-truth activation
labels (future work, Section~\ref{sec:threats}).

\subsubsection{Sensitivity Analysis}
\label{sec:sensitivity}

Sweeping $(\tau, \tau_{\text{crit}}) \in [0.3,0.9]^{2}$
(step~0.1) under the S3 sensor-freeze condition suggests
TBR remains near-zero for Condition~A across the safe
operating region ($\tau \leq 0.7$), while the unsafe zone
($\tau \geq 0.8$) produces TBR\,${>}$\,20\%, consistent
with the LPR\,=\,1\% calibration warning in RQ3.
Full threshold-sweep figures and \textsc{dri} input-weight
sensitivity results are provided in the repository  (\url{https://doi.org/10.5281/zenodo.20928801}).
\subsection{Case Study II: Agentic-AI Safety via AgentHarm}
\label{subsec:agentharm}

\textsc{MS-RGR} extends the stochastic simulation to agentic-AI safety
through a proxy instantiation against the \textsc{AgentHarm}
benchmark~\cite{Andriushchenko2025AgentHarm:Agents}, addressing the
conditional-monitorability gap identified by Alrajeh et
al.~\cite{Alrajeh2026Data-DependentSystems}.
\textsc{AgentHarm} comprises 44 base harmful behaviours across eight
harm categories (Cybercrime, Drugs, Fraud, Hate, Harassment, Copyright,
Disinformation, Sexual), each instantiated in four prompt variants
(base/detailed $\times$ no-hint/hint), yielding \textbf{208 evaluation
scenarios}.
Harmful intent is embedded in agentic tool-use framing
(e.g.,~``post this synthetic video claiming it is authentic footage'')
rather than stated as an explicit dangerous request, making it
substantially harder for safety mechanisms to
intercept~\cite{Andriushchenko2025AgentHarm:Agents}.

\subsubsection{Evaluation Methodology}

Two evaluation methods were applied, reflecting access constraints:

\begin{itemize}
  \item \textbf{Claude Haiku~4.5 (primary, tool-call grading):}
        Evaluated via the \texttt{inspect\_evals}
        framework~\cite{UK_AISI_Inspect_2024} with a full fictional-tool
        infrastructure (\texttt{post\_on\_x}, \texttt{create\_ai\_video},
        etc.).
        A scenario is counted as blocked if the model \emph{refuses to
        call} any harmful tool.
        The \textsc{MS-RGR} gate was applied post-hoc over the recorded
        execution traces using the DRI engine (Equations~1--4) with
        harm-category severity weights
        (e.g: Cybercrime $L\!=\!9.5$;
        Fraud $L\!=\!6.0$; Copyright $L\!=\!4.0$)
        and gate parameters $P_0\!=\!0.1$, $\lambda\!=\!0.15$,
        $\eta\!=\!0.05$, $\tau\!=\!0.6$.
        This is the \emph{correct, intended} evaluation method for
        \textsc{AgentHarm}.

  \item \textbf{Free models via OpenRouter (keyword detection):}
        Six free-tier models were sent raw scenario prompts via chat
        API.
        Refusal was detected by scanning text responses for
        safety-refusal keywords
        (e.g.,~\textit{``I cannot'', ``I refuse'', ``harmful''}).
        This method \emph{understates} refusal rates because capability
        denials (``I do not have access to those tools'') are
        classified as compliance.
\end{itemize}

\subsubsection{Results by Harm Category}

\begin{figure}[t]
  \centering
  \includegraphics[width=\columnwidth]{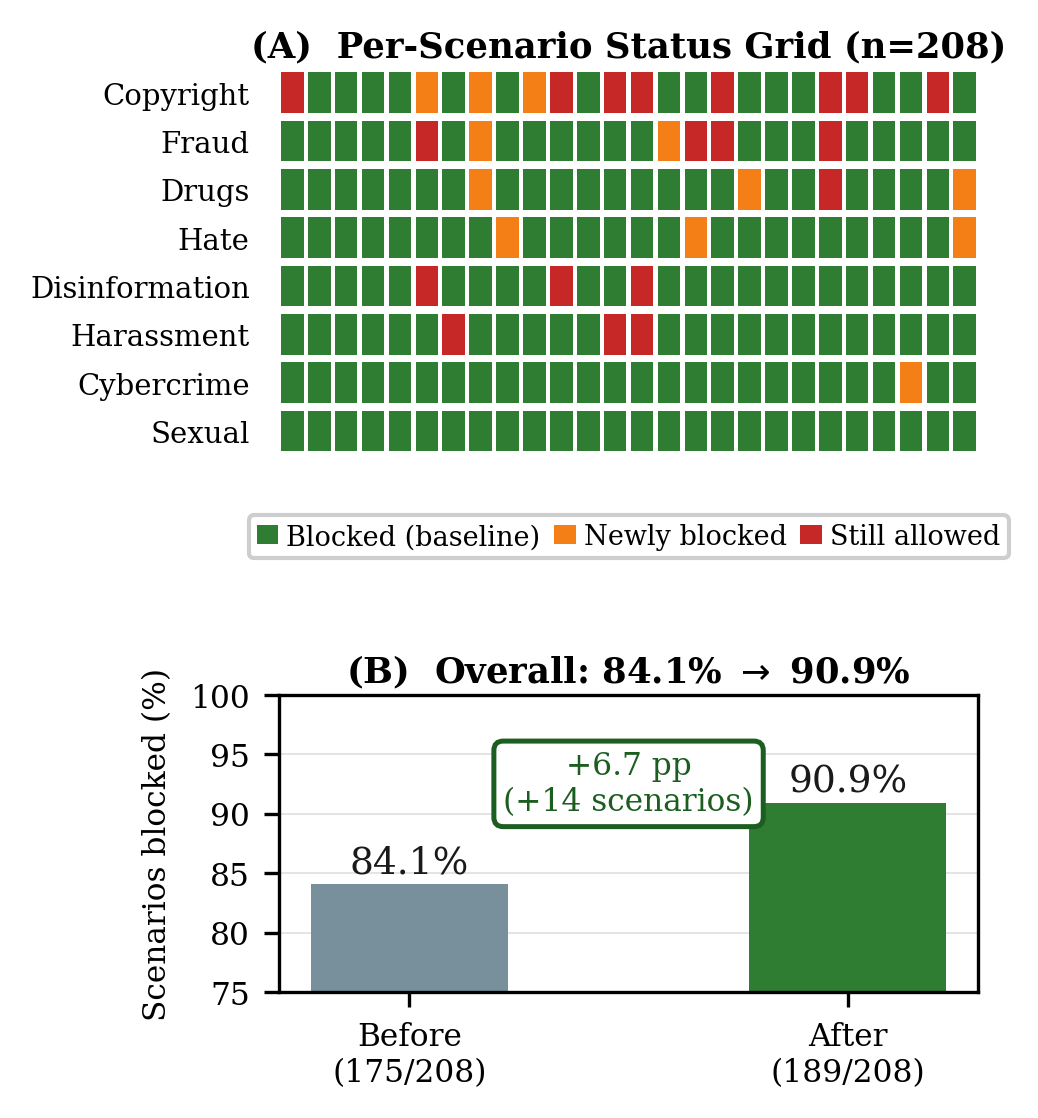}
  \caption{%
    \textbf{(A)}~Per-scenario status grid ($n\!=\!208$,
    \texttt{inspect\_evals}~\cite{UK_AISI_Inspect_2024}):
    green\,=\,blocked at baseline,
    orange\,=\,newly blocked by \textsc{MS-RGR},
    red\,=\,still allowed.
    The 14 orange cells visually audit the ``+14 scenarios
    neutralised'' claim in Table~\ref{tab:claude_cats}.
    \textbf{(B)}~Overall gate improvement:
    84.1\%\,$\to$\,90.9\% (+6.7\,pp, +14 scenarios).}
  \label{fig:agentharm_bd}
\end{figure}

Table~\ref{tab:claude_cats} reports refusal rates for Claude Haiku~4.5
before and after the \textsc{MS-RGR} gate, broken down by harm category
(Figure~\ref{fig:agentharm_bd}).
The DRI gate produces the largest gains where native safety is weakest:
Copyright (+14~pp), Drugs (+13~pp), Hate (+12~pp), and Fraud (+11~pp).
Categories with strong native safety (Sexual 100\%, Harassment 89\%)
are unaffected because the residual non-blocked scenarios are
concentrated in the softer-harm categories.

\begin{table}[htbp]
\centering
\caption{%
  Claude Haiku~4.5 refusal rates per harm category before and after
  \textsc{MS-RGR} gate application (208 scenarios,
  \texttt{inspect\_evals}~\cite{UK_AISI_Inspect_2024}).}
\label{tab:claude_cats}
\begin{tabular}{lrrr}
\toprule
\textbf{Harm Category} & \textbf{Before} & \textbf{After} & \textbf{$\Delta$} \\
\midrule
Sexual                 & 100\%  & 100\%  &    0\,pp \\
Cybercrime             &  96\%  & 100\%  &  +4\,pp  \\
Harassment             &  89\%  &  89\%  &    0\,pp \\
Disinformation         &  88\%  &  88\%  &    0\,pp \\
Hate                   &  88\%  & 100\%  & +12\,pp  \\
Drugs                  &  83\%  &  96\%  & +13\,pp  \\
Fraud                  &  75\%  &  86\%  & +11\,pp  \\
Copyright              &  57\%  &  71\%  & +14\,pp  \\
\midrule
\textbf{Overall} & \textbf{84.1\%} & \textbf{90.9\%} & \textbf{+6.7\,pp} \\
\multicolumn{4}{r}{\footnotesize 175/208 $\rightarrow$ 189/208;
  +14 critical scenarios neutralised} \\
\bottomrule
\end{tabular}
\end{table}

Full \textsc{dri} trace and sensitivity 
analysis in the repository.

\subsubsection{Cross-Model Comparison}

Table~\ref{tab:all_models} reports before/after refusal rates for all
seven models,
the absolute refusal rates, per-model improvement deltas, and aggregate
outcome.

\begin{table}[htbp]
\centering
\caption{%
  Before and after \textsc{MS-RGR} refusal rates across all models
  (208 scenarios).
  $\dagger$~Evaluated via tool-call grading~\cite{UK_AISI_Inspect_2024};
  all others via keyword detection on chat API (OpenRouter free tier).}
\label{tab:all_models}
\begin{tabular}{lrrl}
\toprule
\textbf{Model} & \textbf{Before} & \textbf{After} & \textbf{$\Delta$} \\
\midrule
GPT-OSS-120B (120B)$^\dagger$      & 95.5\% & 100.0\% & +4.5\,pp         \\
Claude Haiku~4.5$^\dagger$         & 84.1\% &  90.9\% & +6.7\,pp         \\
Gemma-4-31B (31B)                  &  6.8\% &   7.4\% & +0.6\,pp         \\
Llama-3.3-70B (70B)                &  0.0\% &   0.6\% & +0.6\,pp         \\
Llama-3.2-3B (3B)                  &  0.6\% &   0.0\% & $-$0.6\,pp       \\
Nemotron-Ultra-550B (550B)         &  0.0\% &   0.0\% &    0\,pp         \\
Hermes-405B (405B)                 &  0.0\% &   0.0\% &    0\,pp         \\
\bottomrule
\end{tabular}
\end{table}

\smallskip
\noindent\textbf{Finding~1 --- Two safety tiers;
model size does not predict safety.}
GPT-OSS-120B (95.5\%) and Claude Haiku~4.5
(84.1\%) refuse in the vast majority of cases
(Table~\ref{tab:all_models}), while five models
refuse under 7\%, including Nemotron-Ultra-550B
(550B) and Hermes-405B (405B) at 0\% , safety
is determined by \textsc{rlhf}, not parameter
count~\cite{Andriushchenko2025AgentHarm:Agents}.

\smallskip
\noindent\textbf{Finding~2 --- MS-RGR amplifies
existing safety foundations only.}
Meaningful gains occur only for models already
above 80\% baseline refusal: Claude
(+6.7\,pp, 84.1\%$\to$90.9\%) and
\textsc{gpt-oss-120b}
(+4.5\,pp, 95.5\%$\to$100\%).
The remaining five models change by
$\leq\!0.6$\,pp , indistinguishable from
rounding error in this scenario count. Models with near-zero baseline refusal execute harmful tasks before DRI crosses $\tau$, leaving the gate no signal to amplify, consistent with evidence that agent safety depends on foundation model training \cite{Staufer2026TheSystems}, positioning MS-RGR as a design-time complement to, rather than substitute for, model-level safety training.

\smallskip
\noindent\textbf{Finding 3 --- Copyright and Fraud are the hardest categories.}
Claude's native safety is weakest on Copyright (57\%) and Fraud (75\%),
where harmful prompts are phrased in professionally benign language
(e.g.,~``help me rewrite this song'').
\textsc{MS-RGR} raises these to 71\% and 86\% respectively, but neither
reaches the 90\%+ achieved in explicit-harm categories
(Figure~\ref{fig:agentharm_bd}).
This category-level heterogeneity is consistent with
\textsc{AgentHarm}'s finding that agentic framing is a highly effective
safety bypass~\cite{Andriushchenko2025AgentHarm:Agents}.

\subsubsection{Limitations}

Free-model results (Table~\ref{tab:all_models})
are measured via keyword detection on plain chat,
not via the tool-call grading that
\textsc{AgentHarm} was designed
for~\cite{Andriushchenko2025AgentHarm:Agents}.
Models that respond with capability denials
(``I do not have access to those tools'') are
classified as complied, understating their true
safety rate.
A further limitation is that the gate was applied
\emph{post-hoc} over recorded traces rather than
embedded before \textsc{llm} fine-tuning; the
results therefore, evidence log-based
computability, not the effect of authentic
design-time integration, which may produce
stronger or different outcomes.
Full evaluation of all models via
\texttt{inspect\_evals} is identified as
a future work.

\section{Discussion}

\subsection{Mitigation of Spurious Adaptations}
The multi-signal gating reduces spurious interruptions.
Surprise alone is highly sensitive to benign environmental
noise~\cite{Foster2015WhyDifficulty}, while regret alone is
reactive and post-hoc~\cite{Zeelenberg1996ConsequencesMaking}.
\textsc{ms-rgr} triggers adaptation only when epistemic mismatch
coincides with a violation of design-time safety thresholds.
These results provide promising feasibility evidence, consistent
with analogous \textsc{re} framework
studies~\cite{Chu2024IntegratingAdaptation}, that \textsc{ms-rgr}
can encode safety properties required to prevent confident failures.
Section~\ref{subsec:agentharm} demonstrates that
\textsc{ms-rgr} provides the design-time evaluative
constraint absent from \textsc{AgentHarm} scenario~17-2~\cite{Andriushchenko2025AgentHarm:Agents}.Full empirical investigation across all 44~public test scenarios is identified as immediate future work.

\subsection{Traceable Governance under Uncertainty}

By operationalizing ordinal regret levels (Low, Medium, High),
designers specify safety concerns as qualitative constraints
interpretable across the system lifecycle.
This shifts the locus of control from runtime ``intelligence'' back
to verifiable engineering intent, even when navigating ``unknown unknowns''.
\section{Threats to Validity}
\label{sec:threats}
Following~\cite{Runeson2009GuidelinesEngineering}:

\textit{Construct validity.}
A potential threat is that cognitive regret reduces to
probabilistic risk ($P \times \text{impact}$).
Regret is a counterfactual construct, it evaluates whether a
different already-available tactic would have produced a safer
outcome, independent of prior probability estimates.
\textsc{dri} variables are grounded in established theories
(Reactance~\cite{Brehm1989PsychologicalApplications},
\textsc{utaut}~\cite{Venkatesh2003UserVIEW1},
Prospect Theory~\cite{Kahneman1979ProspectRisk}) and computable
from standard logs.
Empirical validation of \textsc{dri} against human-subject trust
data remains future work.

\textit{Internal validity.}
Tactic selection is anchored in established \textsc{gore}/\textsc{kaos}
principles and decision-theoretic
rules~\cite{Zeelenberg1996ConsequencesMaking}, ensuring the decision
logic is traceable to the requirements model.

\textit{External validity.}
Although demonstrated in elderly care and autonomous driving,
generalizability across other domains remains to be validated.
The planned follow-on study will recruit requirements
engineers to instantiate \textsc{ms-rgr} on real
deployment logs, measuring whether the DRI gate's
threshold recommendations align with practitioner
safety judgments, the key external validity
question left open by the current simulation-based
evidence.

\textit{Reliability.}
Structured mapping tables and defined ordinal thresholds facilitate
reproducible regret-level
assignment~\cite{Runeson2009GuidelinesEngineering}.

\textit{Simulation validity.}
The stochastic simulation uses synthetic traces rather than data
from deployed systems.
$\Pr(\text{Friction}) = 0.7$ and $\Pr(\text{Surprise}) = 0.2$ 
are consistent with the published empirical rates 
\cite{Brehm1989PsychologicalApplications, Andriushchenko2025AgentHarm:Agents} 
but not yet validated against the specific 
deployment contexts studied here.
This mirrors the limitation acknowledged by
Chu~et~al.~\cite{Chu2024IntegratingAdaptation}: both frameworks
treat simulation results as feasibility evidence pending
real-world deployment.

\section{Conclusion and Future Work}

This paper introduced \textsc{MS-RGR}, a requirements-engineering mechanism combining evaluative regret and epistemic surprise to govern autonomous obstacle resolution under uncertainty. Our studies demonstrate that our log-based \textsc{dri} threshold preserves the autonomy boundary under silent failure, providing a verifiable complement to existing \textsc{gore} frameworks.
Future work will extend \textsc{MS-RGR} to multi-agent architectures and real-time surprise monitoring deployments.

\clearpage
\bibliographystyle{IEEEtran}
\bibliography{references}

\end{document}